\documentclass[aps,preprint,showpacs]{revtex4}

\usepackage{graphicx,epsfig}
\usepackage{amssymb}
\usepackage{float}
\begin{document}

\title{ Absorption section for black holes surrounded by scalar fields}

\author{J.C.  Olvera$^1$ }
\email{jcolvera@fis.cinvestav.mx}

 \affiliation{$^1$ Dpto de F\'isica, Centro de Investigaci\'on y de Estudios Avanzados del I.P.N,
14-740 Apdo, DF, M\'exico}

\begin{abstract}
The absorption section of black holes surrounded by scalar fields is studied with the sinc approximation in the eikonal limit. The effects of the scalar field on the nonlinear Euler-Heisenberg, the magnetic Bardeen and Reissner-Nordstrom BHs are studied in order to obtain information about how the scalar field affects different BHs.

\end{abstract}

\maketitle

\section{Introduction}

One of the most important objects to study in general relativity are black holes because they represent singularities in the spacetime and can be described in an effective way by parameters such as mass, angular momentum and charge \cite{Townsend:1997ku}.   

Black holes (BH) can be analized indirectly by studying fields and matter surrounding it. One way to obtain information about the BH's exterior is to observe the absorption section. For Bardeen BHs the absorption section has been studied in \cite{PhysRevD.90.064001}. This procedure has been used in a variety on contexts, from the study of non linear electrodynamics \cite{Olvera:2020sa}, to gravitational waves \cite{Dolan:2008kf} \cite{Sebastian:2014cka} \cite{Macedo:2015ikq}. 

Absorption sections can be studied from two different approaches, each one depending on the frequency of the absorbed field. The low frequency limit \cite{Higuchi:2001si} and the high frequency limit, commonly known as sinc approximation \cite{Decanini:2011xi}. In the latter case (which will be used here),  the absorption cross section has an oscillatory behaviour in terms of the frequency. When considering the eikonal limit, a dependence in orbit parameters, such as the Lyapunov exponent, appear. 

BHs can be surrounded by different kinds of fields, however, in this sector observables are needed in order to comprehend and for further study of these fields.  Consideration of a surrounding field in spherically symmetric BHs is taken into account by the use of an effective geometry which modifies the test particles trajectory and in consequence affects the absorption sections. The scalar field approach for BHs have been studied in 	\cite{Marsa:1996}, \cite{Carneiro2018} and \cite{Mart:2004}. While in \cite{Kiselev:2003} the author associates the scalar field to quintaessence, in \cite{Visser:2019} a discussion arises whether or not this idea is correct. Another approach is to consider solutions in reduced dimensions which is equivalent to considering a scalar field \cite{Mignemi:1989}. The study of absorption sections for Schwarzchild BH's surrounded by scalar fields has been applied in \cite{Hao:2012}.

The main goal of this  paper is to describe the modifications in the BH's absorption sections produced by considering the scalar field in the way Kiselev did \cite{Kiselev:2003}. We study spherically symmetric BHs in asymptotic form. 

The paper is organized as follows. In Sec. II, a summary of the scalar field metric modification is presented. Sec. III  describes how the absorption section can be obtained by sinc approximation and considering the effects of the scalar field. The BH's absorption sections for the metrics mentioned above are compared with their scalar field contribution in Section IV, while the conclusions are given in the last section.

\section{ effective metric}

Particles around a spherical BH follow a certain trajectory, however, if the BH is surrounded by a field, a spacetime deformation appears produced by this field which modifies the original trajectories.
 	
Following \cite{Kiselev:2003} considering only spherically symmetric metrics with the form 
\begin{equation}\label{eq1}
ds^2=f(r)dt^2-f(r)^{-1}dr^2-r^2(d\theta^2+\sin^2\theta d\phi^2),
\end{equation}
it has been found that the curvature is given by 
\begin{equation}
R=3c\omega_q \frac{1-3\omega_q}{r^{3(\omega_q+1)}},
\end{equation}

where $\omega_q$ represents the scalar field parameter which has the information about what the field represents and $c$ is a positive normalization factor.

The effects on the metric components appear by imposing that the density, i.e. the energy-momentum tensor's trace, to be positive, resulting in the condition $c \omega_q\leq 0$. This implies that $\omega_q<0$.

Adding the scalar field, the effective metric takes the form of equation (\ref{eq1}) by changing $f(r)\rightarrow F(r)$, where 
\begin{equation}\label{eq2}
F(r)=f(r)-\left(\frac{r_q}{r}\right)^{3\omega_q+1}.
\end{equation}

$r_q$ is a normalization parameter, while $\omega_q$ can be associated with the state equation of the scalar field.

The parameter $\omega_q$ must fulfill certain conditions. First, to have de Sitter outer horizons in this model, $\omega_q<-1/3$ must be hold \cite{Hellerman:2001}. By imposing this, the solutions are not asymptotically flat.  Secondly, a lower bound appears given by $\omega>-1$ in order to suppress long range interactions of the scalar field \cite{Carroll:1998}. Combining both conditions results in $-1<\omega_q<-1/3$.

\begin{figure}[H]
\centering
\includegraphics[scale=.65]{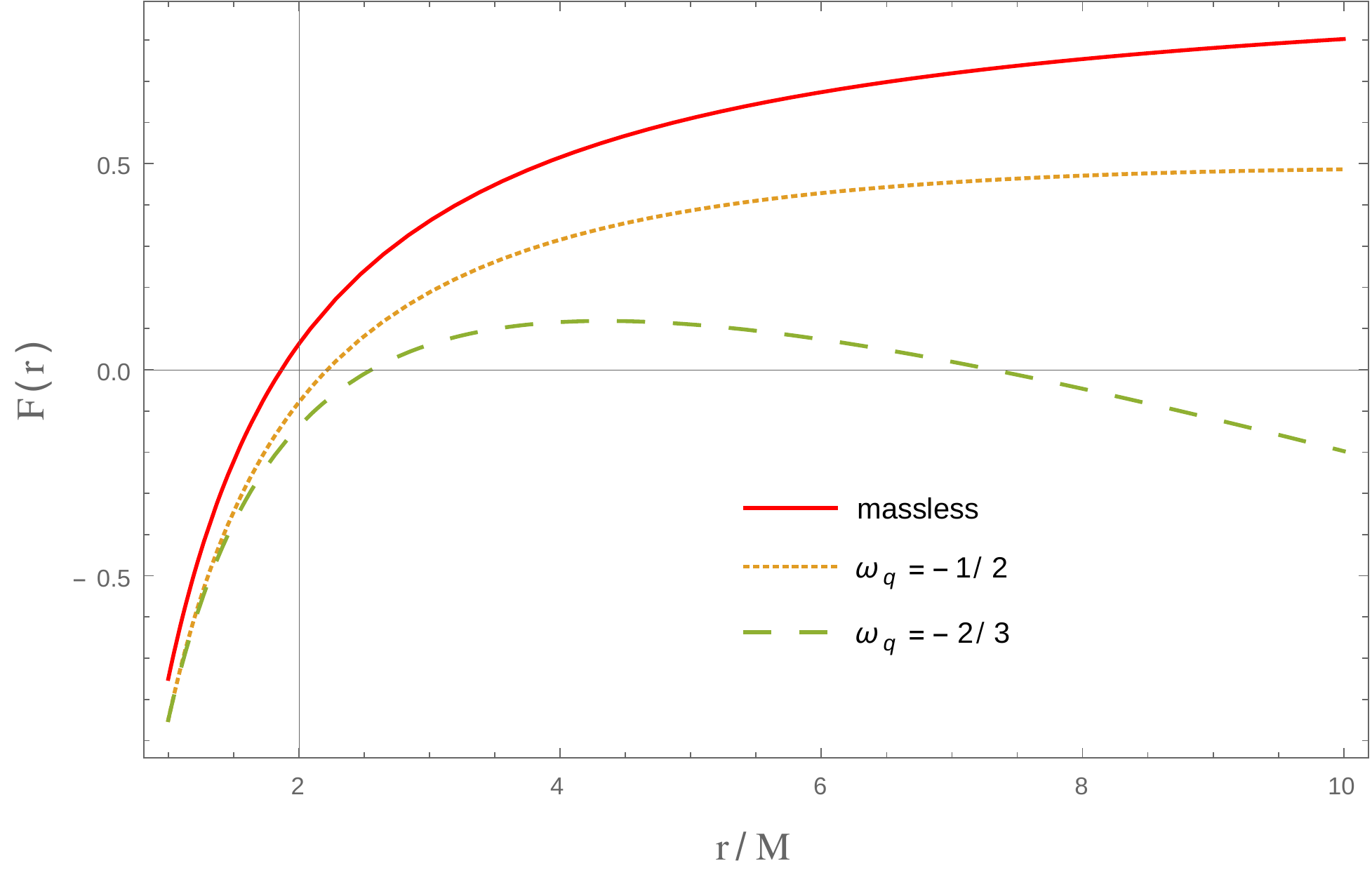}
\caption{Horizons for RN BH, the dotted line represents the scalar field metric. ($\omega_q=-1/2$, $L/M=0.4$, $E=0.1$, $Q=0.5$).}\label{RNF}
\end{figure}

\begin{figure}[H]
\centering
\includegraphics[scale=.65]{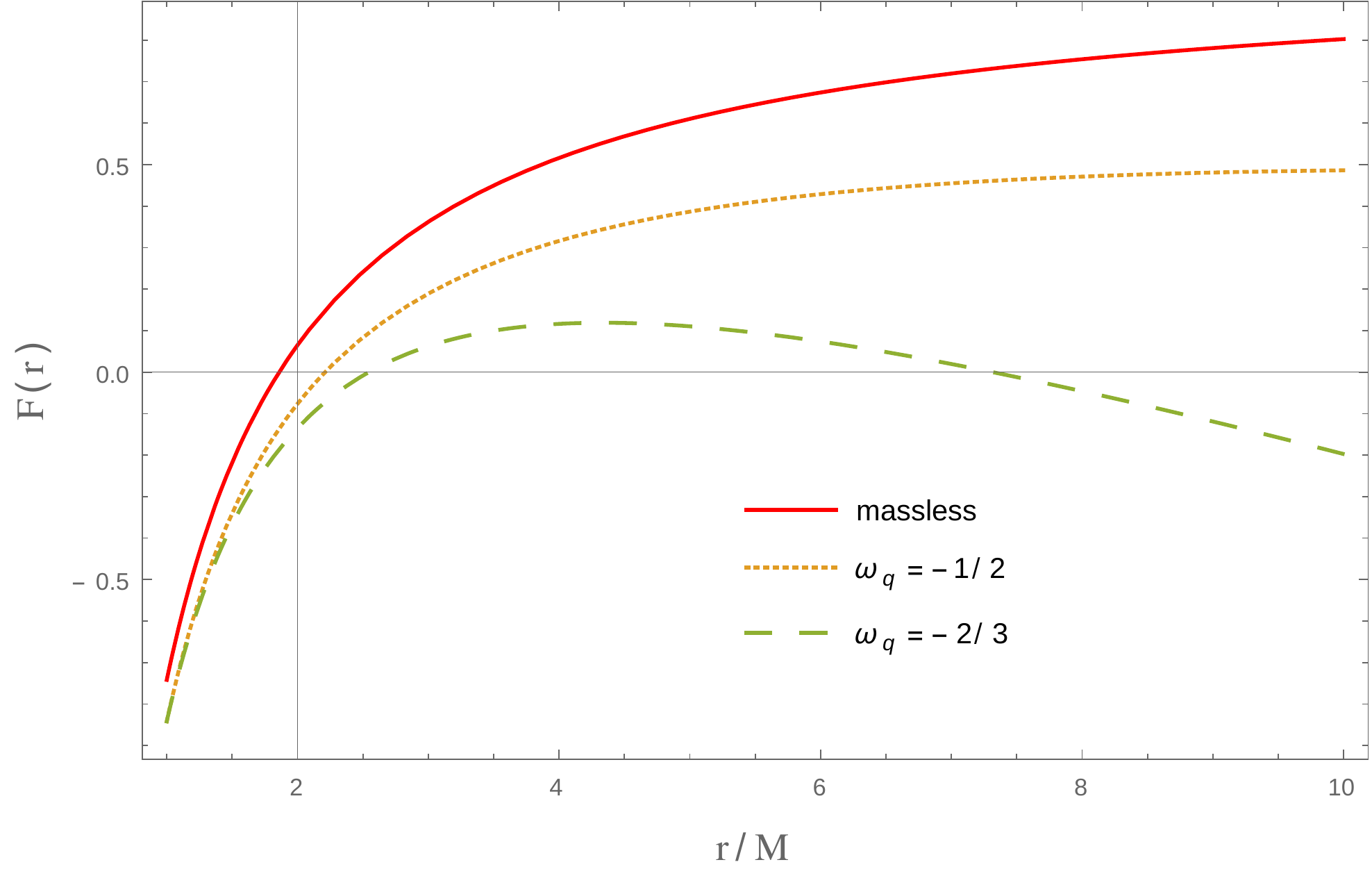}
\caption{Horizons for EH BH, the dotted line represents the scalar field metric. ($\omega_q=-1/2$, $L/M=0.4$, $E=0.1$, $Q=0.5$).}\label{EHF}
\end{figure}

\begin{figure}[H]
\centering
\includegraphics[scale=.7]{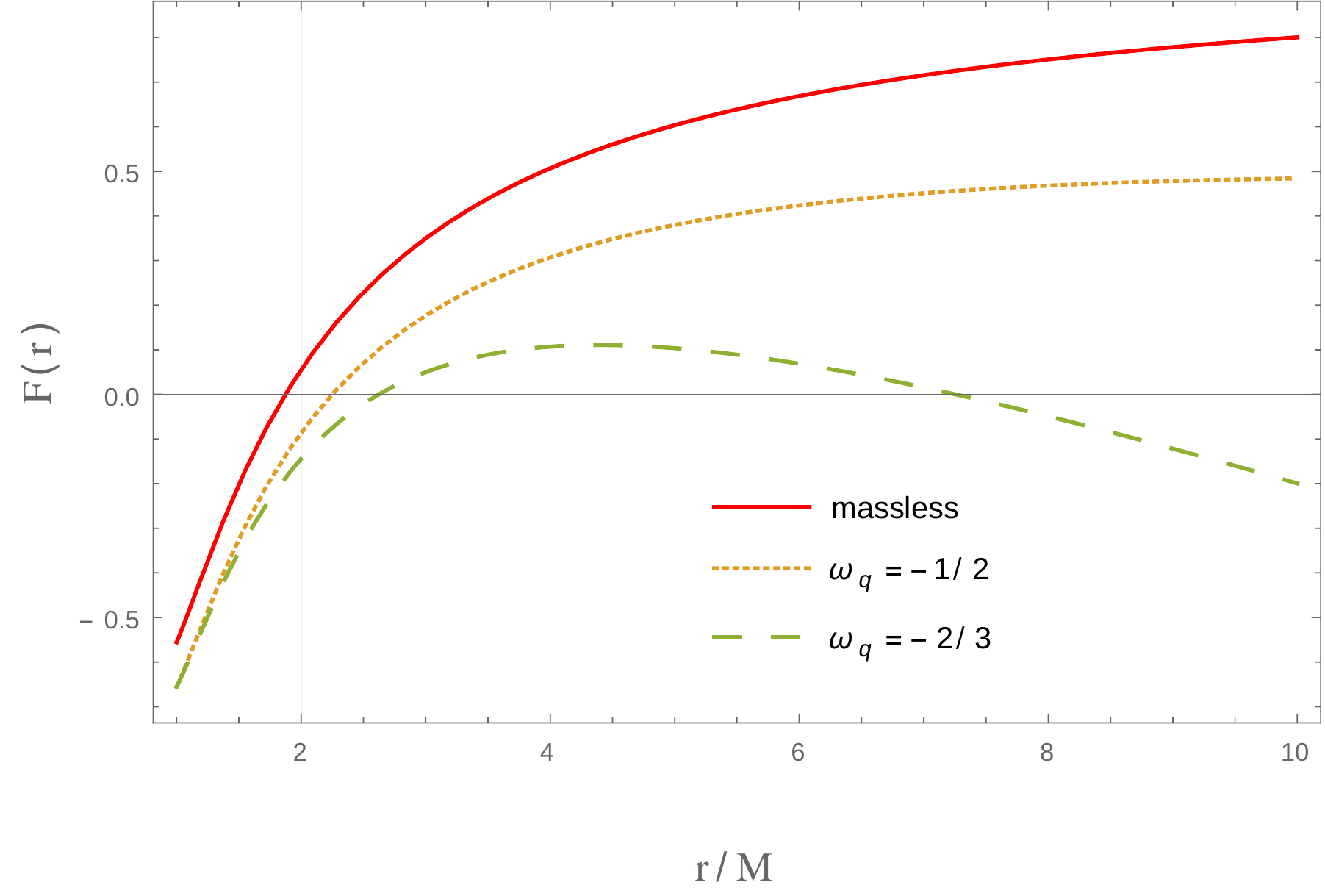}
\caption{Horizons for Bardeen BH, the dotted line represents the scalar field metric. ($\omega_q=-1/2$, $L/M=0.4$, $E=0.1$, $Q=0.5$).}\label{BF}
\end{figure}

For this analysis, it is important to study how the horizons change with respect to the scalar fields. In order to obtain a single horizon, it is necessary to set $\omega_q=-1/2$ for the black holes that are shown in this work, however, setting $\omega_q=-2/3$ produces an inner and outer horizon. The horizons are shown in figures \ref{RNF}, \ref{EHF} and \ref{BF} for RN, EH and Bardeen black holes.

\begin{figure}[H]
\centering
\includegraphics[scale=.65]{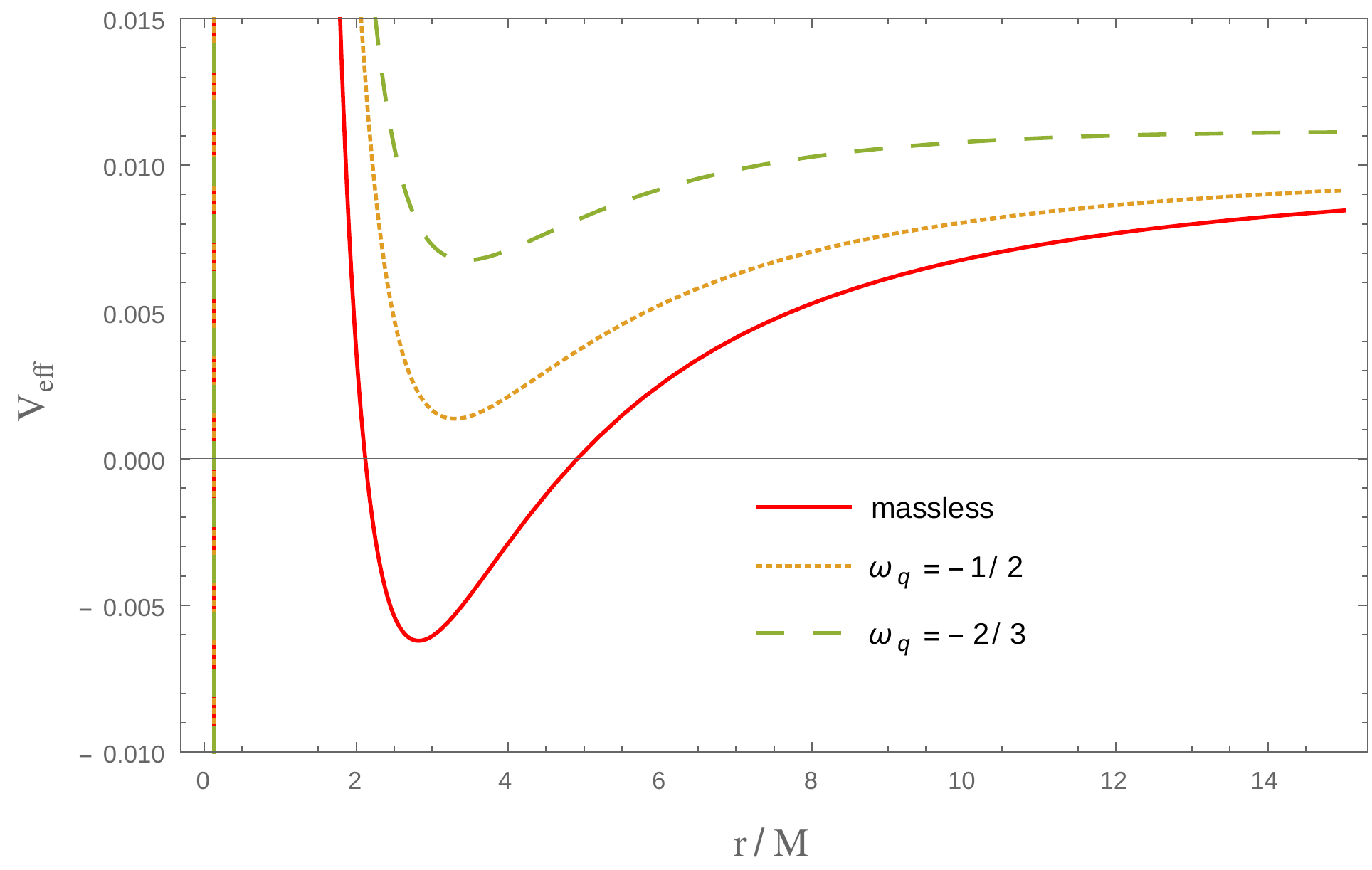}
\caption{Effective potential for RN BH, the dotted line represents the scalar field metric. ($\omega_q=-1/2$, $L/M=0.4$, $E=0.1$, $Q=0.5$).}\label{RNV}
\end{figure}

\begin{figure}[H]
\centering
\includegraphics[scale=.65]{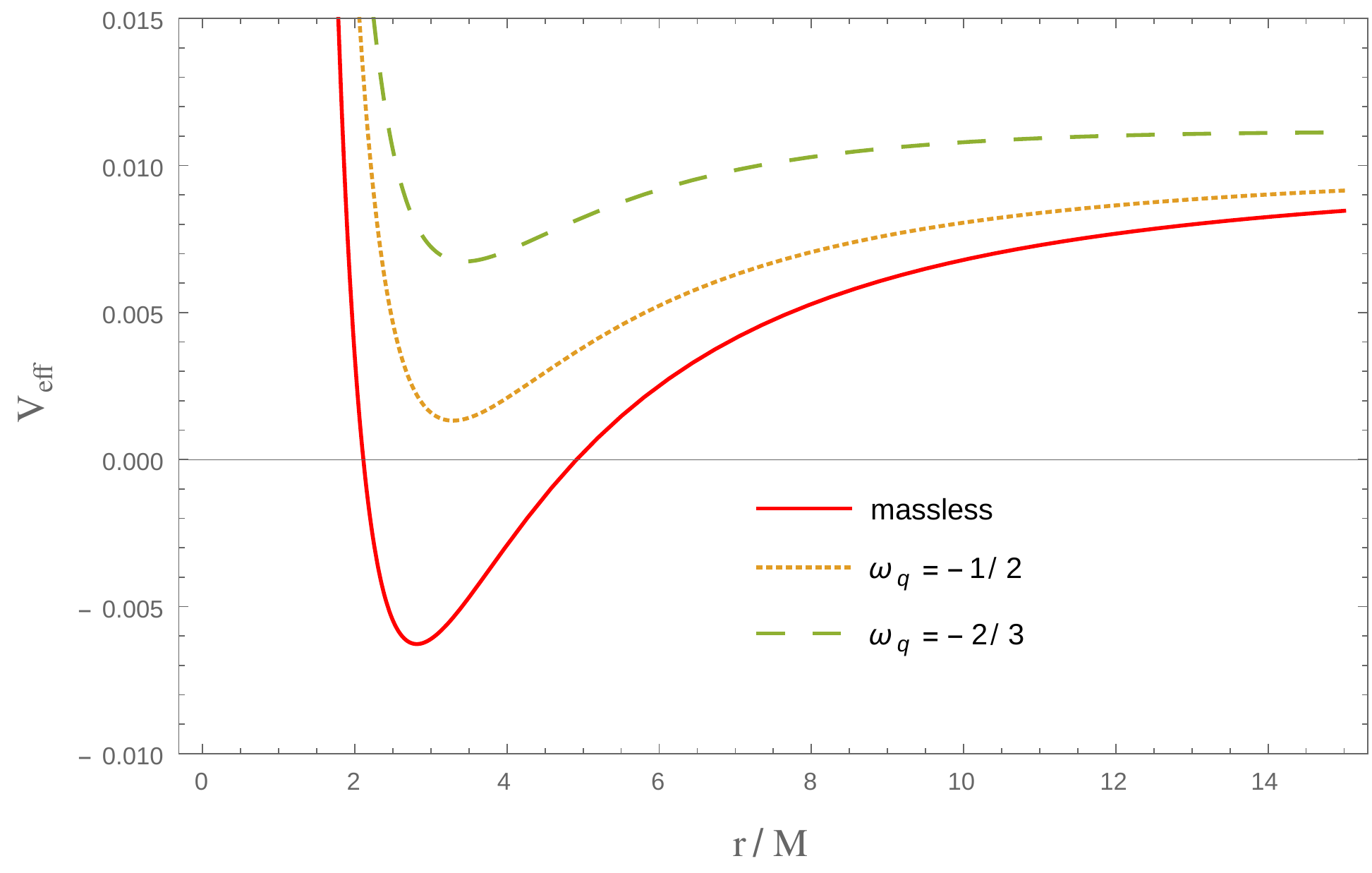}
\caption{Effective potential for EH BH, the dotted line represents the scalar field metric. ($\omega_q=-1/2$, $L/M=0.4$, $E=0.1$, $Q=0.5$).}\label{EHV}
\end{figure}

\begin{figure}[H]
\centering
\includegraphics[scale=.7]{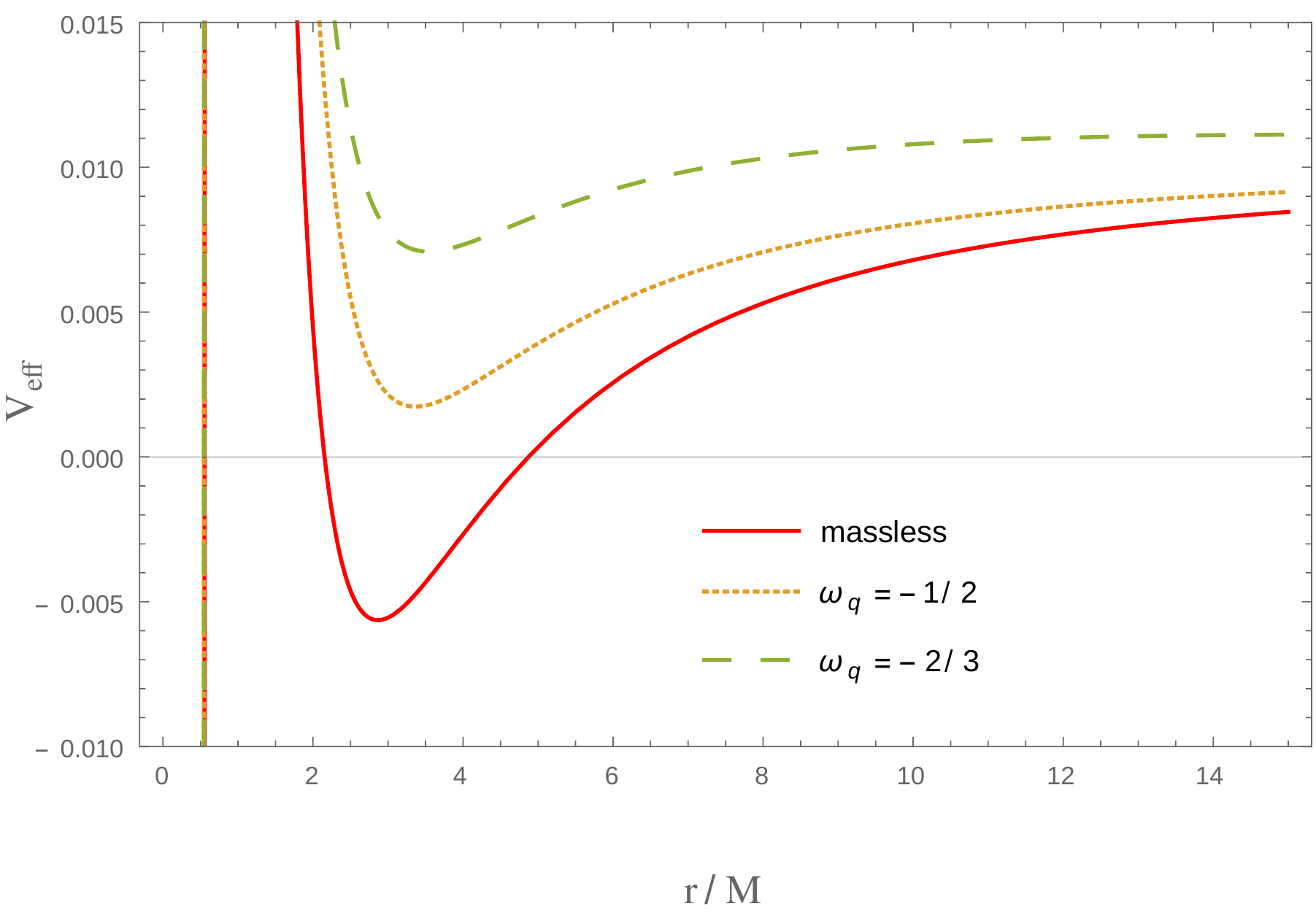}
\caption{Effective potential for Bardeen BH, the dotted line represents the scalar field metric. ($\omega_q=-1/2$, $L/M=0.4$, $E=0.1$, $Q=0.5$).}\label{BV}
\end{figure}

The scalar field affects the effective potentials as well. This can be seen in figures \ref{RNV}, \ref{EHV} and \ref{BV}, from where it can be seen that the circular orbit's radius is going to be higher when in presence of a scalar field. 
\section{Absorption section}

To obtain the absorption section, we study the low frequency limit as well as the mid to high frequencies.

Following \cite{Higuchi:2001si}, if the BHs are far away, the wave solution for 3+1 dimensions is given by 
\begin{equation}
\psi \approx e^{i \omega t} \left[e^{i \omega r \cos\theta}+f_\omega(\Omega) \frac{e^i\omega r}{r^2}\right]
\end{equation}
In the low frequency limit, the absorption section reduces to the BH area.
In the case of  mid to high frequencies the sinc approximation can be used  to analyze the absorption cross section as in \cite{PhysRevD.83.044032}, where we express it in terms of the transmission coefficient.

Thus this absorption section in the eikonal limit is given by ;

\begin{equation}\label{eq3}
\sigma_{abs}=\pi b_{c}^{2} - 4\pi \frac{\lambda b_{c}^{2}}{\omega}e^{-\pi \lambda b_{c}}\sin(2\pi \omega b_{c})
\end{equation}

 where $\lambda$ is the Lyapunov exponent and $b_c$ is the critical impact parameter.
 
 Following the treatment on \cite{Cardos:2009} for circular orbits, the Lyapunov exponent is given by 
 \begin{equation}\label{eq4}
 \lambda^2=\frac{f(r_c)}{2r_c^2}\left[2f(r_c)-r_c^2 f''(r_c)\right].
 \end{equation}
 
 Because the quasinormal modes are approximated in the eikonal limit by 
 \begin{equation}
\Omega_{n}= b^{-1}_{c}- \imath(n+1/2)|\lambda|,
\end{equation}
 the impact parameter takes the form 
 \begin{equation}
 b_c=\sqrt{\frac{r_c^2}{f(r_c)}},
 \end{equation}
with $r_c$ representing the circular orbit radius, which can be obtained from the condition for circular orbits,
\begin{equation}
\left[2f(r)-rf(r)\right]|_{r=r_c}=0.
\end{equation}

With the methodology explained above absorption sections can be computed for spherically symmetric BHs just by knowing the BH metric. Some examples are analized in the following section.
\section{Examples}

We calculate the absorption sections for black holes using the effective metric given by equation (\ref{eq2}) for the BHs mentioned before, as mentioned in \cite{Ghaderi:2017}. Then, we compare the differences with the case when the effective metric is not considered.

\subsection{Bardeen black hole}

 The Bardeen BH is usually interpreted as a nonlinear magnetic monopole \cite{AyonBeato:2000zs} with mass $M$ and magnetic charge $q_{m}=g$, which is also self-gravitating.  By using the asymptotic form of the Barden BH,

\begin{equation}\label{Bardeenmetric1}
F_{Bardeen}(r)=1-\frac{2M}{r}+\frac{3 M g^2}{r^3}-\frac{c}{r^{3\omega_q+1}}.
\end{equation}

In \cite{Macedo:2017szm} \cite{Huang:2014nka},  absorption cross sections were studied for a massless scalar wave in Bardeen black hole, also in \cite{Macedo:2015qma}, the cross section was computed.

 To carry out the study of Bardeen's absorption section, we have used $r\rightarrow rM$  and $g\rightarrow \frac{4M}{3\sqrt{3}}g$.

\begin{figure}[H]
\centering
\includegraphics[scale=.7]{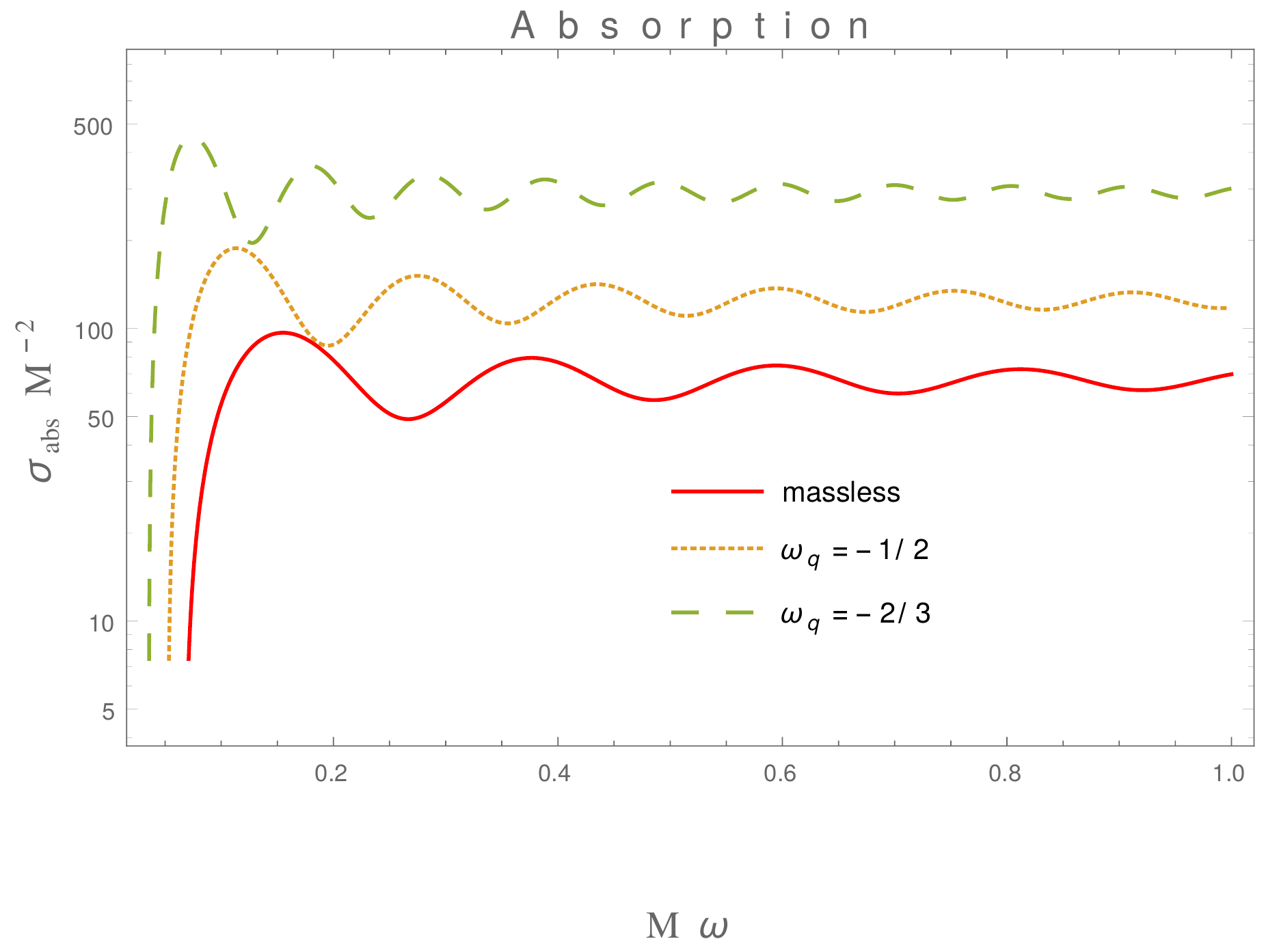}
\caption{Absorption section for Bardeen BH, the dotted line represents the scalar field metric. $Q=0.8$}\label{F1}
\end{figure}

When studying the absorption section in the sinc approximation, it is possible to observe how the effect of the scalar field increases the absorption section with respect to the original one, as can be appreciated in figure  \ref{F1}. 

\subsection{RN  black hole}

The RN black hole, which corresponds to a charged BH, is represented by the metric function in the line element of the form (\ref{eq2}) as:
\begin{equation}
F_{RN}(r)=\frac{2M}{r}+\frac{Q^2}{r^2}-\frac{c}{r^{3\omega_q+1}},
\end{equation}
with $Q$ and $M$ being the charge and mass of the BH.

\begin{figure}[H]
\centering
\includegraphics[scale=.65]{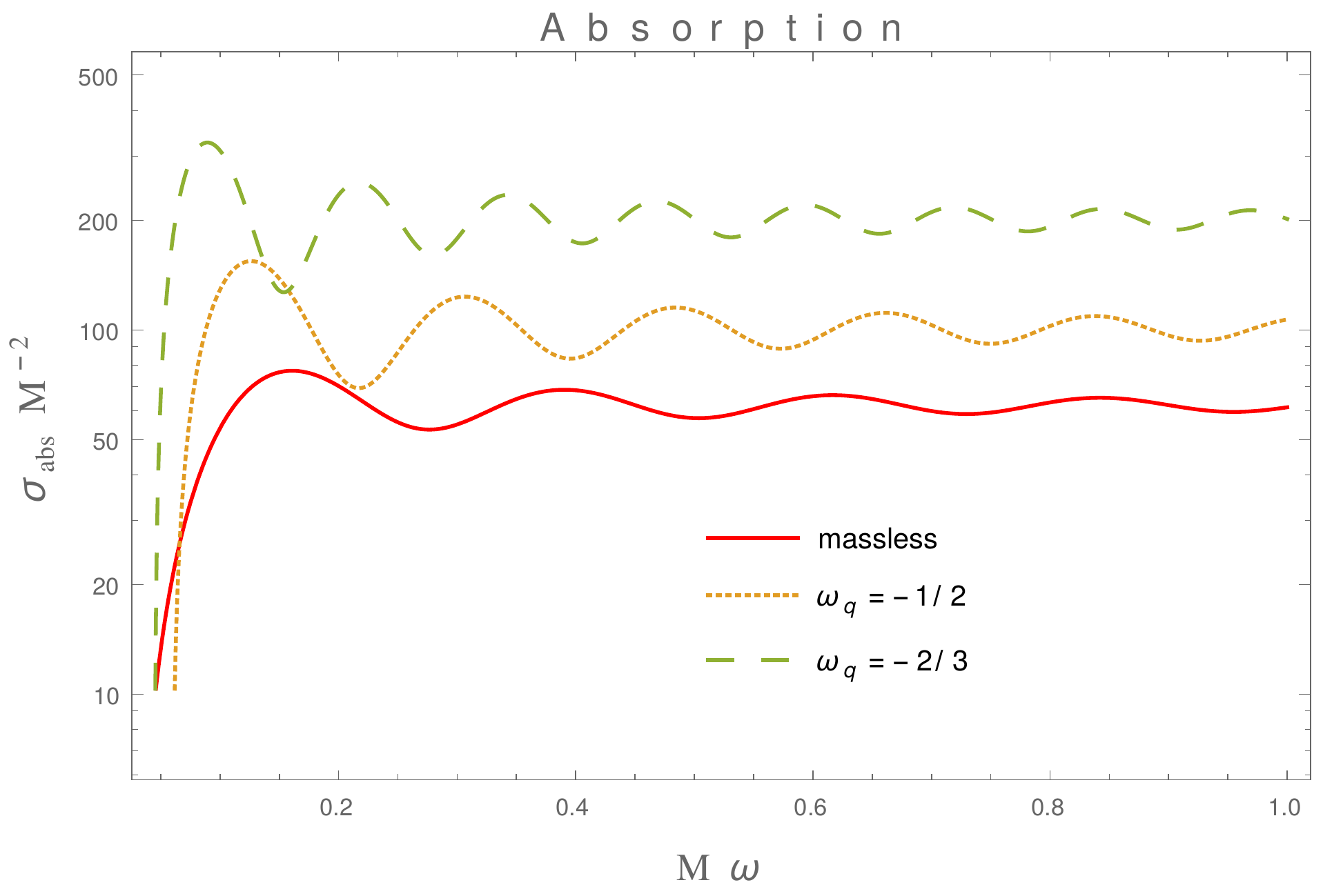}
\caption{Absorption section for RN BH, the dotted line represents the scalar field metric. $\omega_q=-1/2$, $Q=0.8$}\label{F2}
\end{figure}

For the RN absorption section (see figure \ref{F2}) it  is possible to observe that the case of considering the scalar field increases in a similar way to the behaviour from the Bardeen absorption section.

\subsection{Euler-Heisenberg Black hole}

The Euler-Heisenberg black hole is obtained from one-loop quantum corrections in QED \cite{Yajima:2000kw}. The lagrangian associated to the effective action is given by ;

\begin{equation}\label{EHlagr}
L(F)= F(1-aF),
\end{equation}

with $a=he^{2}/(360\pi^{2}m_{e}^{2})$, where $h$, $e$, and $m_{e}$ are the Planck constant, electron charge, and electron
mass respectively.

The corresponding metric  element in  (\ref{eq2}) given by;

\begin{equation}
F_{EH}(r)=1-\frac{2M}{r}+\frac{q^{2}}{r^{2}}-\frac{2}{5}a\frac{q^{4}}{r^{6}}-\frac{c}{r^{3\omega_q+1}},
\end{equation}

where $M$ is the mass parameter and $q$ is the magnetic charge.  One or two horizons may occur: if  $(M/q)^{2}\leq 24/25$ a single horizon exists but for  $(M/q)^{2}> 24/25$, a second and a third horizons occur. 

As in the cases shown before, we consider the following $r\rightarrow M r$, $a\rightarrow \frac{8}{27}q^2$ and $q\rightarrow \frac{5 M}{\sqrt{24}}q$.

\begin{figure}[H]
\centering
\includegraphics[scale=.65]{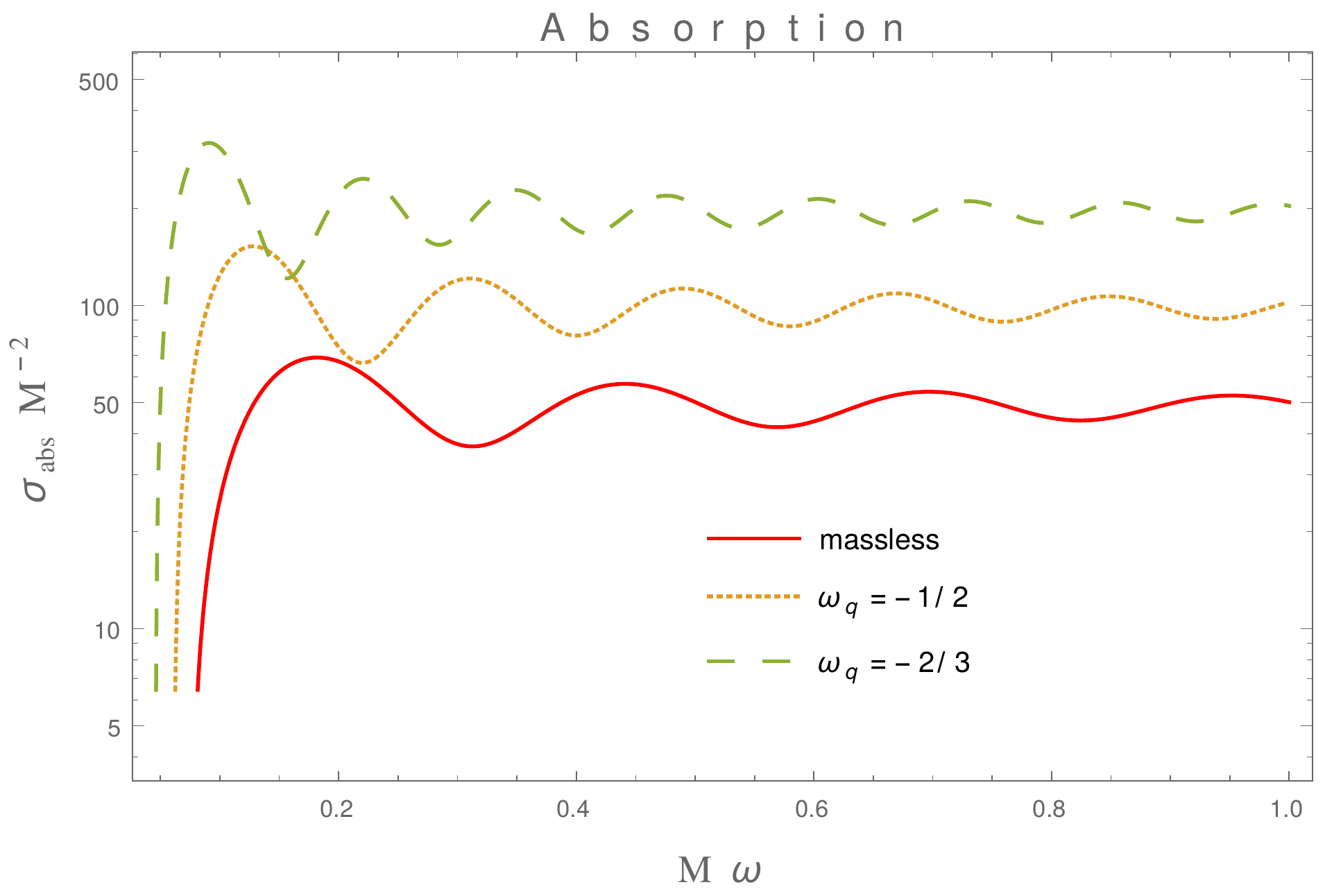}
\caption{Absorption section for EH BH, the dotted line represents the scalar field metric. $\omega_q=-1/2$, $Q=0.8$}\label{F3}
\end{figure}

We studied the absorption section (see figure \ref{F3}), when we consider the case with a scalar field, significant differences are present in comparison with the original case.

\section{Conclusions}
In this work, it is firstly shown how the absorption sections are modified by considering the effective metric generated by the action of an external scalar field for null geodesics. The sinc approximation is used, so that the absorption section is given in function of the Lyapunov exponent and the impact parameter for a null circular unstable geodesic, which  was also modified by considering the effective metric.

The first aspect that can be seen from the figures, is that by including the scalar field, an increase in the absorption section appears, as well as an oscillation frequency shift, being this more notable in the Bardeen BH case and less in RN.

A possible interpretation for this result is that besides the attraction produced by the black hole, the surrounding scalar field produces an spacetime deformation as well, increasing the effective area of interaction.

When considering lower values for $\omega_q$ the geometric absorption section, as well as the frequency for the oscillatory section, increases significantly, which can be associated to the change in the circular orbit's radius. 

In order to understand how to distinguish BHs and BHs surrounded by scalar fields further investigation is needed such as other observables like the scattering sections, which is considered for future work.

\bibliographystyle{unsrt}

\bibliography{bibliografia}

\end{document}